\documentclass[letterpaper, 10 pt, conference]{ieeeconf}  

\IEEEoverridecommandlockouts                              

\usepackage{multirow}
\usepackage{tabularx}
\usepackage{adjustbox}
\usepackage{makecell}
\usepackage{subfig}
\usepackage{comment}
\usepackage{amsmath,amsfonts,amssymb}

\title{\LARGE \bf
A Novel Speech Intelligibility Enhancement Model based on Canonical Correlation and Deep Learning}

\author{Tassadaq Hussain$^{1}$, Muhammad Diyan$^{1}$, Mandar Gogate$^{1}$, Kia Dashtipour$^{1}$,  \\
Ahsan Adeel$^{2}$, Yu Tsao$^{3}$, and Amir Hussain$^{1}$ 
\thanks{$^{1}$are with the School of Computing ,
        Edinburgh Napier University, UK
        {\tt\small {t.hussain, m.diyan, k.dashtipour, m.gogate, a.hussain}@napier.ac.uk}}%
\thanks{$^{2}$Ahsan Adeel is with the Department of Electrical Engineering, University of Wolverhamption, UK. 
        {\tt\small a.adeel@wlv.ac.uk}}%
\thanks{$^{3}$Yu Tsao is with the Research Center for Information Technology, Academia Sinica, Taipei, Taiwan.  
        {\tt\small yu.tsao@citi.sinica.edu.tw}}%
}

\begin{document}

\maketitle
\thispagestyle{empty}
\pagestyle{empty}

\begin{abstract}
Current deep learning (DL) based approaches to speech intelligibility enhancement in noisy environments are often trained to minimise the feature distance between noise-free speech and enhanced speech signals. Despite improving the speech quality, such approaches do not deliver required levels of speech intelligibility in everyday noisy environments . Intelligibility-oriented (I-O) loss functions have recently been developed to train DL approaches for robust speech enhancement. Here, we formulate, for the first time, a novel canonical correlation based I-O loss function to more effectively train DL algorithms. Specifically, we present a canonical-correlation based short-time objective intelligibility (CC-STOI) cost function  to train a fully convolutional neural network (FCN) model. We carry out comparative simulation experiments to show that our CC-STOI based speech enhancement framework outperforms state-of-the-art DL models trained with conventional distance-based and STOI-based loss functions,  using objective and subjective evaluation measures for case of both unseen speakers and noises. Ongoing future work is evaluating the proposed approach for design of robust hearing-assistive technology.

\end{abstract}

\section{INTRODUCTION}
Speech enhancement (SE) aims to improve the intelligibility and quality of speech signals in real-time environments that have been distorted by additive and convolutive sounds. In recent years, non-linear spectral mapping-based approaches have shown excellent performance for SE tasks. For example, a deep denoising autoencoder (DDAE) \cite{lu2013speech} and deep neural network (DNN) \cite{xu2014regression}-based frameworks were initially proposed for SE and demonstrated excellent results in suppressing noise components and improving quality and intelligibility of the estimated speech signal. In addition to conventional DNN structures, convolutional neural network (CNN) and long short-term memory (LSTM)-based frameworks were also  used in an attempt to further improve generalisation performance in seen and unseen noisy conditions. In \cite{pandey2019tcnn}, a real-time SE model was proposed by training a CNN framework in an encoder-decoder style. The authors in \cite{fu2017raw} proposed an end-to-end SE framework based on fully convolutional neural network (FCN) to recover the enhanced speech waveform. Later, by replacing a conventional loss function with an objective evaluation-based cost function, an improved FCN-based SE framework was proposed to enhance speech perception in noisy environments \cite{fu2018end}.

Recently, Reddy et al. \cite{reddy2022performance} used a straightforward CNN model for SE to achieve a slightly better performance compared to more complicated algorithms. In order to train the model, different types of noise were inputted and PESQ was used as an evaluation metric. It is important to note that despite reducing the parameters, there was no major improvement observed in the performance of the model. Hussain et al. \cite{hussain2022novel} integrated temporal attentive-pooling (TAP) into the CNN approach for the SE task. The convolutional layer was used to extract the local information of audio signals and  RNN was used to characterize temporal contextual information. The infant cry dataset was used to evaluate performance and experimental findings indicated that the approach can effectively reduce noise level from infant cry signals.

\begin{figure}[!t]
    \centering
    \includegraphics[height = 1.2in, width=0.5\textwidth]{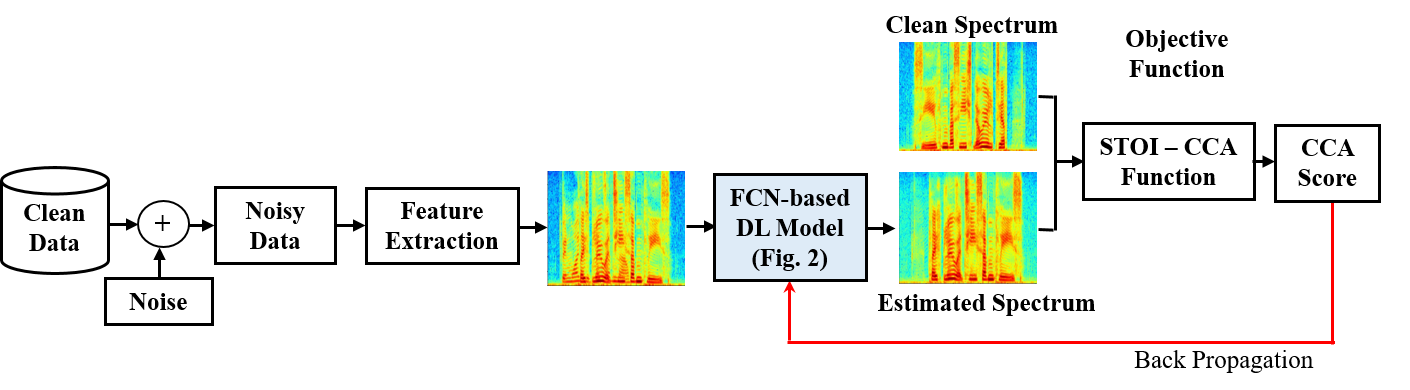}
    \caption{Proposed CC-STOI-based SE model}
    \label{fig:block}
\end{figure}

Despite the outstanding performance of DL-based SE models, distance-based loss functions, like mean squared error (MSE) and mean absolute error (MAE), are commonly utilised to optimise the parameters of such systems. These cost functions, on the other hand, are not based on human auditory perception and may not be appropriate for speech-related applications. We believe that employing human perception-based evaluation measures to optimise such systems could result in more optimal task-related outcomes. Researchers have lately begun using a range of objective evaluation measures based on human auditory perception to optimise DL-based systems. Perceptual evaluation of speech quality (PESQ) \cite{rix2001perceptual} and short-time objective intelligibility (STOI) \cite{taal2011algorithm} are among those two widely used metrics for measuring speech quality and intelligibility. For example, a number of intelligibility-oriented (I-O) STOI-metric based DL algorithms have recently been proposed and proven to be useful for SE. The authors in \cite{fu2018end} employed a STOI metric as an objective function to optimize an FCN model for SE. The proposed I-O STOI-based system proved to be effective and demonstrated better results than a standard MSE-based for SE task due to increased consistency between the training and evaluation target. 

In this paper, we propose a first of its kind canonical-correlation based short-time objective intelligibility (CC-STOI) based cost function to quantify and optimise  correlations between noisy and clean speech signals. 
We focus on the magnitude spectra of noisy and clean speech utterances as we process the signal frame by frame in the frequency domain. Next, we utilise an FCN to learn the spectral mapping function using a CC-STOI function. Apart from formulating a CC-STOI loss function, we evaluate the performance of CC-STOI-based system against two loss functions, namely STOI and MSE. All SE frameworks are trained and evaluated on the benchmark GRID corpus \cite{cooke2006audio} using two background noise scenarios, such as speech and non-speech background noises, generated by synthetically mixing clean utterances with speech/non-speech background noises to form noisy data. 

Experimental results demonstrate that the DL-based framework when optimized using a CC-STOI function, can achieve significant SE performance improvement over both MSE and STOI-based SE methods under different noise and SNR conditions in terms of standardized evaluation metrics: namely, PESQ, STOI, and speech distortion index (SDI) \cite{le2019sdr}.

\begin{figure}[!t]
    \centering
    \hspace*{-0.3cm}
    \includegraphics[height = 1.5in, width=0.52\textwidth]{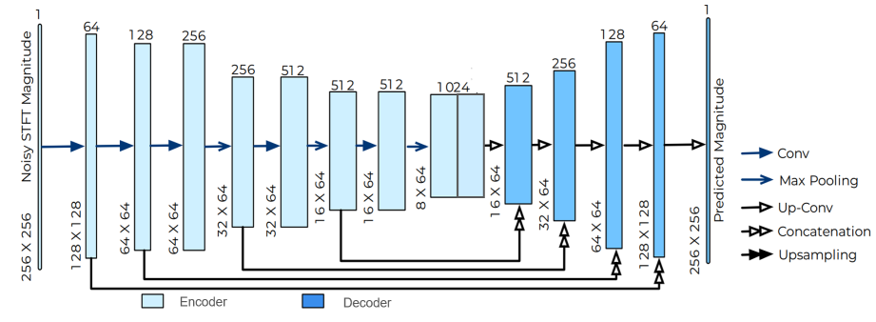}
    \caption{Proposed FCN-based framework for CC-STOI-based SE}
    \label{fig:framework}
\end{figure}

\section{Methodology}
Current DL-based SE frameworks are normally trained to minimize the MSE between estimated and target magnitude spectrums. As a result, we start with an FCN-based U-Net model ~\cite{ronneberger2015u} for SE that has been trained with an MSE loss function. The MSE loss function between the noisy and estimated magnitude spectrum can be computed as:

\begin{equation}
 \mathcal{L}_{MSE} = min (\frac{1}{N} \sum_{n=1}^{N} \|\hat{Y}_n - Y_n\|_2)
\end{equation} 
where $\textit{\^Y}_N$ and $\textit{Y}_N$ is the estimated and corresponding target magnitude spectrum,  $\textit{N}$ is the number of speech frames, and $ \| \cdot \|_2$ represents L2-normalization.

Similar to \cite{hussain2021towards}, we adopted a deep FCN-based U-Net architecture for SE utilizing a CC-STOI loss function to estimate the enhanced magnitude spectrum. The proposed CC-STOI-based SE model is shown in Figure 1, and the FCN-based U-Net framework is shown in Figure 2. Our goal is to evaluate how incorporating CCA into the STOI loss function affects the quality and intelligibility of the estimated speech signal when compared to standard MSE and STOI loss functions.

\subsection{Short-time Objective Intelligibility based on Canonical Correlation Analysis}

To train a frequency domain I-O SE model, we utilized a modified STOI proposed in \cite{hussain2021towards} as an objective function. The STOI function takes the clean and estimated speech signals as input and computes the score by: i) removing the silent frames from clean and estimated speech signals, ii) applying the short-time Fourier transform (STFT), iii) Estimating the envelope of clean and noisy speech using one-third octave-band analysis of the STFT frames, iv) Normalizing and clipping to compensate for global level differences and stabilisation of the STOI evaluation, and v) measuring intelligibility. To optimise the correlation between the two spectral envelopes, we replaced the correlation coefficient of standard STOI with the CCA. The correlation coefficient between the two speech signal is estimated using the equations below to optimise the CC-STOI function.

 \begin{equation}
 \textit{d}_{i,j} = \frac {(y_{i,j} - \mu_{y_{i,j}})^\intercal  (\hat{y}_{i,j} - \mu_{y_{i,j}})^\intercal } {\|y_{i,j} - \mu_{y_{i,j}}\|_2  \|\hat{y}_{i,j} - \mu_{\hat{y}_{i,j}}\|_2}
\end{equation} 
 where $\textit{y}$ and $\textit{\^y}$ represents the spectral envelope of the target and estimated speech signals, $\mu_{y_{i,j}}$ and $\mu_{\hat{y}_{i,j}}$ are the corresponding sample mean vectors, and $\| \cdot \|_2$ represents the L2-normalization. The final CC-STOI is the average of the intelligibility measure over all bands and frames.
 
 \begin{equation}
 \textit{d}_{CC-STOI} = \frac {1} {I(M - N + 1)} \sum_{i=1}^{I} \sum_{j=1}^{J} \textit{d}_{i,j}
\end{equation} 
 where $I$ = 15 represents the number of one-third octave bands and $M - N + 1$ presents the total number of short-time temporal envelope vectors. Please see \cite{taal2011algorithm} for a more complete description of each step. Because the CC-STOI calculation is differentiable, it may be used directly as the objective function to optimise the SE model.

 \begin{equation}
  \mathcal{L}_{CC-STOI} = - \frac {1} {M} \sum_{m=1}^{M} \textit{d}_{CC-STOI}(\hat{Y}_m, Y_m)
\end{equation} 

The CC-STOI score between the estimated and clean magnitude spectra of audio utterances is measured using $\textit{d}_{CC-STOI}(\hat{Y}_m, Y_m)$. Unlike MSE, where the goal is to reduce the distance, we want to improve speech intelligibility by maximizing the CC-STOI score.

\subsection{Proposed SE Framework}
This section presents the proposed FCN-based U-Net framework for SE as shown in Fig. 2. The network architecture for feature extraction and the speech resynthesis pipeline are discussed in detail below.

\subsubsection{Audio feature extraction}
For SE, the speech feature extraction step employs a U-Net~\cite{ronneberger2015u} style network incorporated with an audio SE-modified encoder and  decoder block. The magnitude of noisy speech STFT of dimension $F \times T$ where $F$ and $T$ are the frequency and time dimensions of the spectrogram, is fed into the network. Moreover,  input is provided for two convolutional layers consisting of 4 and 2 strides to reduce time-frequency until it reaches 64. The reduced features are  processed via three convolutional blocks, each with convolution layers that have a filter size of three and stride of one, following which the frequency pooling layer is used to reduce the size of the frequency dimension by two. It is important to note that amidst the execution of convolutional blocks, spatial dimension is  preserved. When noisy spectrogram is provided as an input, the proposed model estimates the clean spectrogram, and when using an inverse STFT, the predicted magnitude is blended with the noisy phase to produce enhanced speech. 

The decoder is made up of three convolutional blocks each with two upsampling layers to upsample the dimension by two. The upsampling layer is followed by convolutional layers with a filter size of three. The acoustic features are then fed into two  transposed convolutional layers with a filter size of four and a stride of two to upsample the TF dimension until it equals the input dimension. To keep the output in the 0 to 1 range, we utilise a sigmoid activation function. The predicted mask is then multiplied by the input spectrogram to produce the output masked spectrogram. When the noisy STFT magnitude is provided as input, the proposed model estimates the magnitude of clean STFT. Using an inverse STFT, the predicted magnitude is combined with the noisy phase to generate enhanced speech.

\section{Experiments and Results}

\subsection{Experimental Setup}
A small vocabulary GRID corpus \cite{cooke2006audio} is utilised to train our proposed I-O SE model and assess how the CC-STOI loss function influences entire SE efficiency. The dataset contained a video recordings of 34 male and female speakers, each with 1000 utterances lasting roughly three-seconds. In this paper, we only utilized the audio data to assess the performance of our proposed framework. The audio data was initially recorded at 48 kHz sampling rate which subsequently resampled to 16 kHz for processing. In this paper, we randomly selected ten speakers for the training set, two speakers for validation, and three speakers for the test sets. To evaluate the performance of the proposed I-O SE framework, we employed three objective evaluation measures to measure the quality, intelligibility, and speech to noise distortion index, i.e., PESQ, STOI, and SDI. 

\subsection{STOI vs CC-STOI}
To assess the correlation between CC-STOI and modified STOI \cite{hussain2021towards} that accounts a 16 kHz frequency domain signal while ignoring down sampling and silent frame removal steps, we first plot a scatter plot between CC-STOI and modified STOI scores. The scatter plot is used to observe the relationship between CC-STOI and modified STOI scores. Figures 3(a) and 3(b) show scatter plots for CC-STOI and modified STOI scores computed between clean and noisy utterances, and clean and enhanced utterances. The scatter plots show that the CC-STOI score correlates well with the modified STOI score and can be used for training DL models.

\begin{figure}[!t]
        \centering
    \subfloat[\centering ]{{\includegraphics[width=0.25\textwidth]{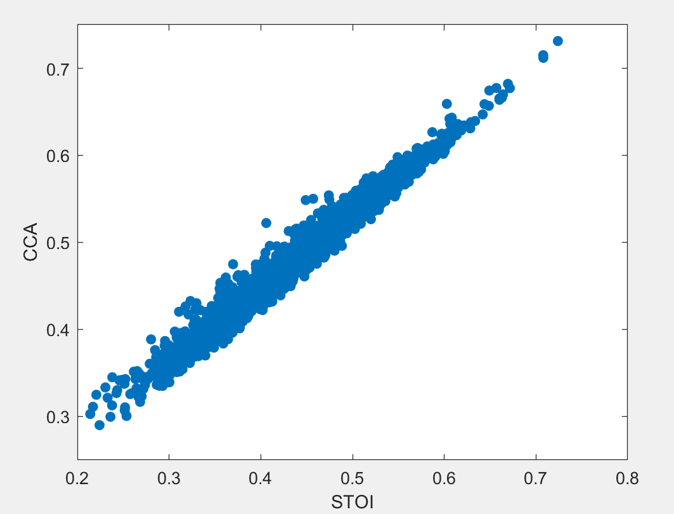} }}%
    \subfloat[\centering ]{{\includegraphics[width=0.25\textwidth]{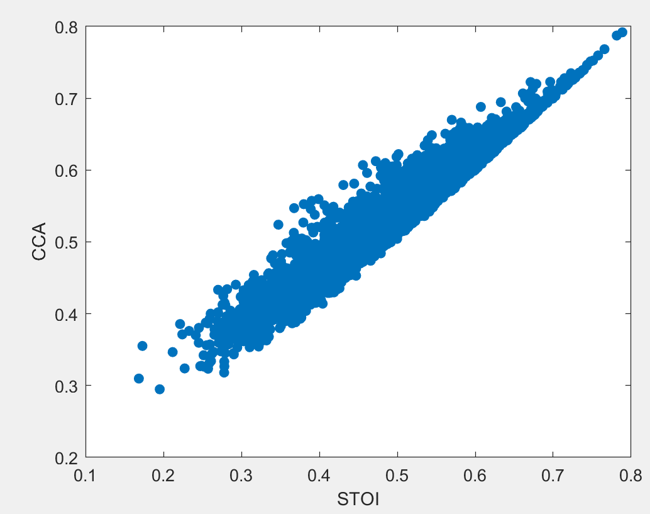} }}%
    \caption{Scatter plot between (a) CC-STOI and STOI scores for clean and noisy utterances and (b) CC-STOI and STOI scores for clean and enhanced utterances}%
    \label{fig:example}%
\end{figure}

\subsection{Speech vs Non-speech Background Noises}

We generated two sets of noisy data, i.e., speech and non-speech, to evaluate the performance of CC-STOI. Initially, we contaminated clean utterances with speech noises ranging from 0 to 20 dB SNR, to generate speech noisy data. For more challenging task, we later contaminated clean utterances with four real-world background noises provided by CHiME-3 corpus \cite{barker2015third} between -12 and 9 dB SNR at a step of 3 dB, namely Bus, Cafeteria, Pedestrian, and Street (termed Bus, Caf, Ped, and Str), to generate non-speech noisy data.

The performance of DL SE frameworks trained using three loss functions for speech and non-speech background noises is compared in Tables I and II. Tables I and II show that DL SE frameworks, when trained with different loss functions, significantly improves the overall quality and intelligibility of noisy speech signal. It can also be seen that DL-based framework with three loss functions also provides less distortion ratio compared to original noisy speech. In summary, the frameworks optimised with the three loss functions were successful for SE. However, we found that, in comparison to SE frameworks trained with the $\mathcal{L}_{MSE}$ and $\mathcal{L}_{STOI}$ loss functions, DL framework trained with $\mathcal{L}_{CC-STOI}$ performed better in terms of PESQ, STOI, and SDI scores, respectively.

\begin{table}[t]
\setlength\tabcolsep{5pt}
\caption{\scshape Performance comparison of DL models using standardised objective evaluation metrics using speech noisy data under speaker-independent conditions.}
\centering
\begin{tabular}{c|c|ccc|c}
\hline
\hline
\multirow{2}{*}{\textbf{Framework}}  & \multirow{1}{*}{\textbf{Loss}} & \multicolumn{3}{c|}{\textbf{Objective Evaluation Metrics}} & \multirow{2.2}{*}{\textbf{Avg.}} \\ \cline{3-5}

  & \multirow{1}{*}{\textbf{Function}} & \multicolumn{1}{c}{PESQ} & \multicolumn{1}{c}{STOI} & \multicolumn{1}{c|}{SDI} \\ \cline{2-5}
 
\hline
\hline
\textbf{Noisy} & -- & 2.442 
& 0.875 & 0.237 & 1.184 \\ \hline
\ & MSE & 2.746 & 0.893 & 0.193  & 1.277 \\
\textbf{U-Net} & STOI & 2.802 & 0.912 & 0.212  & 1.308 \\
 & CC-STOI & 2.831 & 0.912 & 0.225 & 1.322 \\  

\hline
\hline
\end{tabular}
     \end{table}
\begin{table}[t]
\setlength\tabcolsep{5pt}
\caption{\scshape Performance comparison of DL models using standardised objective evaluation metrics using non-speech noisy data under speaker-independent conditions.}
\centering
\begin{tabular}{c|c|ccc|c}
\hline
\hline
\multirow{2}{*}{\textbf{Framework}}  & \multirow{1}{*}{\textbf{Loss}} & \multicolumn{3}{c|}{\textbf{Objective Evaluation Metrics}} & \multirow{2.2}{*}{\textbf{Avg.}} \\ \cline{3-5}

  & \multirow{1}{*}{\textbf{Function}} & \multicolumn{1}{c}{PESQ} & \multicolumn{1}{c}{STOI} & \multicolumn{1}{c|}{SDI}  \\ \cline{2-5}
 
\hline
\hline
\textbf{Noisy} & -- & 1.868 
& 0.647 & 2.036 & 1.517 \\ \hline
\ & MSE & 1.975 & 0.704 & 2.013  & 1.564 \\
\textbf{U-Net} & STOI & 2.107 & 0.792 & 2.055  & 1.651 \\
 & CC-STOI & 2.155 & 0.801 & 2.064  & 1.673 \\  

\hline
\hline
\end{tabular}
     \end{table}
In terms of MOS with CHiME3 noises, Fig.4 shows the results of three distinct speech augmentation approaches. In comparison to MSE loss and STOI loss, the experimental results demonstrate that the CC-STOI performed better. Furthermore, as compared to MSE and STOI losses, the CC-STOI performed better in lower SNRs. The subjects were normal-hearing listeners and were asked to report the results in terms of mean opinion score (MOS). Fig.4 depicts the MOS score of the GRID corpus for non-speech CHiME-3 background noises at {-9, 9 dB} SNR at a step of 3 dB using three loss functions. The experimental results show that $\mathcal{L}_{CC-STOI}$ achieved better speech perception performance in terms of noise suppression when compared with $\mathcal{L}_{STOI}$ and $\mathcal{L}_{MSE}$ especially under low SNR conditions.  

\section{Conclusion}
In this study, we proposed a new canonical-correlation based I-O technique to improve the training and generalisation performance of conventional DL-based SE systems by using a more effective intelligibility evaluation metric as an alternate cost function. In particular, a customised canonical-correlation-based solution has been developed that utilised a canonical-correlation based version of the modified STOI loss function to train the DL SE framework. Our results show that utilising canonical correlation as part of a STOI-based DL system improves not only the intelligibility but also the quality of the output estimated signal by exhibiting less distortion. Overall, the CC-STOI loss function performed well and produced better results for a variety of SE assessment criteria, revealing the potential of modified STOI for optimization of frequency-domain SE applications. We plan to expand on this work by evaluating the performance of CC-STOI based SE for more complex real-world audio-visual datasets. This ongoing future work will pave the way for design of future of multi-modal hearing-assistive technologies.

\begin{figure}[!t]
    \centering
    \hspace*{-0.3cm}
    \includegraphics[height = 2in, width=0.47\textwidth]{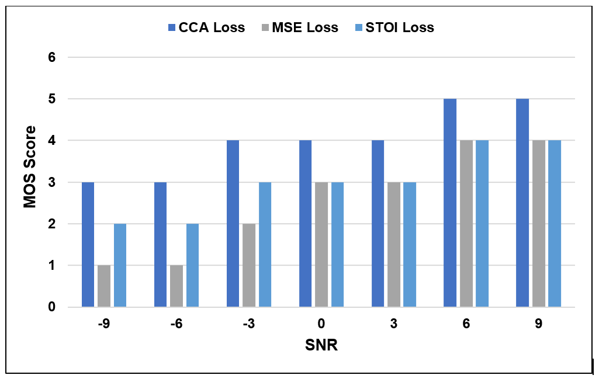}
    \caption{Average subjective listening scores of CC-STOI, STOI, and MSE loss functions for \{-9, 9 dB\} SNR of non-speech GRID corpus.}
    \label{fig:framework2}
\end{figure}

 \bibliographystyle{IEEEtran}
 \bibliography{mybib}

\end{document}